\documentclass[prl,reprint,twocolumn,preprintnumbers,amsmath,amssymb,showpacs,footinbib,superscriptaddress]{revtex4-1}

\usepackage[titletoc,toc,title]{appendix}
\usepackage{braket}
\usepackage[final]{feynmp}
\usepackage{ifpdf}
\usepackage{comment}
\usepackage{amsmath}
\usepackage{mathrsfs}
\usepackage{color}
\usepackage{bm}
\usepackage[font=small]{subcaption}
\usepackage[font=small]{caption}

\graphicspath{{./Figures/}}

\makeatletter
\def\endfmffile{%
	\fmfcmd{\p@rcent\space the end.^^J%
			end.^^J%
			endinput;}%
	\if@fmfio
		\immediate\closeout\@outfmf
	\fi
	\ifnum\pdfshellescape=\@ne
		\immediate\write18{mpost \thefmffile}%
	\fi}
\makeatother

\usepackage[demo]{graphicx}
\usepackage{dcolumn}
\usepackage{bm}
\usepackage{hyperref}

\newcommand{\beq}{\begin{equation}}
\newcommand{\eeq}{\end{equation}}

\def\8{\infty}
\def\oh{\frac{1}{2}}

\def\undertext#1{\vtop{\hbox{#1}\kern 1pt \hrule}}

\def\tr{\hbox{tr}\,}

\def\pbyp#1#2{\frac{\partial#1}{\partial#2}}

\def\be{\begin{equation}}
\def\ee{\end{equation}}
\def\bea{\begin{eqnarray} & &}
\def\eea{\end{eqnarray}}

\def\rf#1{(\ref{#1})}

\def\cZ {{\cal Z}}
\def\p {{\bf p}}

\begin{document}


\title{Spectral form factors of  unconventional superconductors}

\author{Sankalp Gaur}
\author{Victor Gurarie}
\affiliation{Department of Physics and Center for Theory of Quantum Matter, University of Colorado, Boulder, Colorado 80309, USA}

\begin{abstract}
We show that spectral form factors of unconventional gapped superconductors have singularities occurring periodically in time. These are the superconductors
whose gap function vanishes somewhere in momentum space (Brillouin zone) but whose fermionic excitation spectrum is fully gapped. Many, although not all, of these superconductors are topologically nontrivial. In contrast, conventional fully gapped superconductors have 
featureless spectral form factors which are analytic in time. Some gapless superconductors may also have singularities in their spectral form factors, but they are not
as ubiquitous and their
appearance may depend on the details of the interactions among fermionic particles which form the superconductor and on the underlying lattice where the particles move. 
This work builds on the prior publication \cite{Gaur2022}  where Loschmidt echo of topological superconductors, related but not identical to spectral form
factors, was studied. It follows that spectral form factors could be used as a test of the structure of the superconducting gap functions.   \end{abstract}

\maketitle



Spectral form factor is a way to characterize the energy spectrum of quantum systems. It is defined as a trace of the evolution operator 
\be {\cal Z} = \tr e^{-i \hat H t} = \sum_n e^{-i E_n t},
\ee
and can be written as a sum over all the energy levels $E_n$ of a quantum system. It is closely related to the thermal partition function of a quantum system, coinciding
with its formal analytic continuation to the complex values of temperature. 

Fourier transform of the absolute value square of the spectral form factor produces the correlation between energy levels of the system
\be \int dt \, e^{i \omega t} \left| {\cal Z} \right|^2 = 2 \pi \sum_{nm} \delta \left(\omega-E_n+E_m \right) .
\ee
Behavior of the spectral form factors in quantum many body systems attracted some attention recently \cite{Joshi2022,Cotler2017}. 

Here we show that spectral form factors in unconventional gapped superconductors have singularities which occur periodically in time, while their conventional counterparts have
featureless spectral form-factors. More precisely, consider the gap function $\Delta({\bf p})$, where ${\bf p}$ is the (quasi)momentum. In unconventional and especially in topological superconductors it cannot be nonzero everywhere.  Rather it vanishes at points, lines or surfaces in the momentum space or in the Brillouin zone. Consider the spectrum of Bogoliubov quasiparticles $E({\bf p})$ for those values ${\bf p}$ where $\Delta({\bf p})=0$. Suppose $E_-$ is the minimum of $E_{\bf p}$ for all such ${\bf p}$. If $\Delta({\bf p})$ vanishes at a single point, then we obviously  cannot  minimize $E_{\bf p}$ and instead take $E_-$ to be equal to $E({\bf p})$ calculated at this  point. 
Then the singularities occur at times $t_n=\pi (1+2n)/(2 E_-)$ where $n$ is an arbitrary integer.  We show that the nature of the singularities depends on the dimensionality of space and
is given by
\be 
\pbyp{\ln \cal Z} {t}\sim \ln \left| t - t_n \right|
\ee
in two dimensional superconductors and
\be 
\pbyp{\ln \cal Z} {t}\sim \sqrt{\left| t - t_n \right|}
\ee
in three dimensional superconductors. 

For superconductors whose underlying fermionic particle move on a lattice, the singularities could also occur at $t_n=\pi(1+2n)/(2E_+)$, where $E_+$ is the maximum of the 
excitation spectrum computed where $\Delta({\bf p})=0$ (if $E_+$ is different from $E_-$). The existence of these singularities is not as ubiquitous as that of the ones associated 
with $E_-$. 


Let us now demonstrate this by  first studying the example of the two dimensional $p_x+ip_y$ (often abbreviated as $p+ip$) superconductor. Specifically, we have in 
mind attractively interacting identical fermions with the Hamiltonian \cite{Gurarie2007}
\be \label{eq:pwave} H = \sum_{\bf p} \epsilon(p) \hat a^\dagger_{\bf p} \hat a_{\bf p} - \frac{\lambda}{V} \sum_{{\bf p}, {\bf k}, {\bf q}} {\bf k} \cdot {\bf q} \, \hat a^\dagger_{\frac{\bf p} 2 +{\bf k} }
\hat a^\dagger_{\frac{\bf p}2  -{\bf k}} \hat a_{\frac {\bf p} 2 -{\bf q} }
\hat a_{\frac {\bf p} 2 +{\bf q}}. 
\ee
Here \be \label{eq:kinetic} \epsilon(p) = \frac{p^2}{2m} - \mu
\ee is the kinetic energy of these interacting spinless fermions, $\lambda$ is the interaction constant and $V$ is the volume of the system. 
These fermions are known to form a  $p_x+i p_y$ paired fermionic superfluid, which for brevity we will refer to as a $p$-wave superconductor. 
It is a class D superconductor \cite{Ryu2010} which is topological if $\mu>0$ and has a gap as long as $\mu \not =0$. 
The Bogoliubov-de-Gennes (BdG) Hamiltonian of this superconductor takes the following standard form
\be \label{eq:ham} \hat H =  \sum_{{\bf p}, \ p_y>0} \left( \begin{matrix} \hat a^\dagger_{\bf p} & \hat a_{-{\bf p}} \end{matrix} \right) 
\left( \begin{matrix} \epsilon(p) & \Delta({\bf p})  \cr \bar \Delta ({\bf p}) & -\epsilon(p) \end{matrix} \right)
\left( \begin{matrix} \hat a_{\bf p} \cr \hat a^\dagger_{-\bf p} \end{matrix} \right). 
\ee
Here 
 $\Delta({\bf p}) = \left( p_x+i p_y \right)  \Delta_p$ and $\bar \Delta({\bf p})=  \left(p_x -i p_y \right) \bar \Delta_p$ are the gap functions. $\Delta_p$ and $\bar \Delta_p$ are  the magnitudes of the gap functions (the subscript $p$ emphasizes that these are $p$-wave gap functions). To avoid double counting, the summation over ${\bf p}$ is restricted to $p_y>0$. Below all the sums over ${\bf p}$ for $p$-wave superconductors will be restricted in this way. 

Let us use the BdG Hamiltonian to calculate the spectral form factor.
To do that, we diagonalize the BdG Hamiltonian for each ${\bf p}$. Its eigenvalues $\omega_{\pm}(p)$ are 
\be \omega_{\pm}(p) = \pm E(p), \ee
where \be  \label{eq:pwavesp} E(p) =   \sqrt{\epsilon(p)^2 + p^2 \bar \Delta_p \Delta_p }.
\ee
Therefore the trace of its evolution operator is
\be \label{eq:sp} \cZ = \prod_{\bf p} S_{\bf p}, \ S_{\bf p} =  e^{-i t E(p)} + e^{i t E(p)}  =  2 \cos \left( t E(p) \right). 
\ee
Before proceeding to study $\cZ$, let us briefly discuss its analytic properties. Each factor $S_\p$ is obviously an analytic function of time $t$. However, if $S_\p$ vanishes for some
values of $\bf p$ at some critical time $t=t_c$ with all $S_\p$ remaining nonzero if $t$ deviates from $t_c$, this could make $\cZ$ nonanalytic at $t_c$ (we postpone the discussion whether $S_{\bf p}$ can indeed behave in this way until later).
Indeed, suppose $S_{\bf p}$ vanishes at $t=t_c$ if ${\bf p}={\bf p}_c$. Quite generally we should expect that in the vicinity of ${ \bf p}={\bf p}_c$ and $t=t_c$, $S_{\bf p}$ has the following expansion 
\be \label{eq:expa} S_{\bf p} \approx C \left(t-t_c + \alpha \left| {\bf p}-{\bf p}_c\right|^2 \right),
\ee
where $\alpha$ and $C$ are some complex constants (we will see later that, even though it may not be obvious right now, the factors $S_{\bf p}$ are generally complex valued). This immediately leads to
\be \pbyp{ \ln \cZ}{t} =\sum_{\bf p} \pbyp{\ln S_\p}{t} \approx  \sum_{\bf p} \frac{1}{t-t_c + \alpha \left| {\bf p}-{\bf p}_c \right|^2}.
\ee
On the right hand side above the approximate expression for $S_\p$ valid with $\p$ in the vicinity of $\p_c$ and $t-t_c$ small is substituted. 
The sum above is obviously a singular function of time at $t=t_c$, with the details of the singularity dependent on the dimensionality of space and on whether ${\bf p}_c$ is zero or nonzero. This makes ${\ln \cZ}$ as well as ${\cal Z}$ itself a nonanalytic function of time at $t=t_c$ (note an obvious similarity between the thermal free energy and
$\ln \cZ$ introduced above). 

Let us now go back to Eq.~\rf{eq:sp}. For a Hamiltonian \rf{eq:ham} with given $\Delta_p$, $\bar \Delta_p$, and $\epsilon(p)$, Eq.~\rf{eq:sp} gives the answer for its spectral form factor. However, in a superconductor,
$\Delta_p$ and $\bar \Delta_p$ are not fixed beforehand but must be determined self-consistently, by matching the Hamiltonian \rf{eq:pwave} with the BdG Hamiltonian
\rf{eq:ham}. To understand how to do it, let us recall that to calculate thermal partition function
$\tr \exp(-\hat H/(k_B T))$, 
we must determine $\Delta_p$ and $\bar \Delta_p$ by solving the gap equation. In a $p$-wave superconductor, it takes the form
\be \label{eq:gapim} \frac 1 {V} \sum_{\bf p} \frac{p^2 \, \tanh \left[ \frac{E(p)}{2 k_B T} \right]} {E(p)} = \frac{1}{\lambda},
\ee
where $T$ is the temperature and $k_B$ is the Boltzmann constant. This equation is solved for the product $\bar \Delta_p \Delta_p$
which enters $E(p)$. The solution to this equation can be used for example to calculate the thermal partition function of the superconductor. In order to adapt this to calculating the spectral form factor, we replace
$1/(k_B T) \rightarrow i t$, with the result
\be  \label{eq:gapreal} \frac  i {V}  \sum_{\bf p}  \frac{p^2 \, \tan \left[ \frac {t E(p)}{2}  \right]} {E(p)} = \frac{1}{\lambda}.
\ee
This should be understood as an equation to determine $\bar \Delta_p \Delta_p$, which should then be substituted into Eq.~\rf{eq:sp}. In principle, there could be many solutions of the equation \rf{eq:gapreal}. To find the one we should use we should first find the solution of equation \rf{eq:gapim} for the temperatures $T$ where the solution $\bar \Delta_p \Delta_p$ is nonzero, and then analytically continue it to the imaginary values of $T$.


Now observe that this equation predicts that $\bar \Delta_p \Delta_p$ must not be real. Indeed, if it is real, the left hand side of this equation is necessarily imaginary, while the right hand side is real. Similar situation occurs in evaluation of the Loschmidt echo where one also finds \cite{Gaur2022}  that $\bar \Delta$ and $\Delta$ are not complex conjugates of each other. 
With $\bar \Delta_p \Delta_p$ being  complex, $E(p)$ is also generally complex.

As a result, the factors $\cos \left( t E(p) \right)$ generally do not vanish at any $t$. 
The exception to that is $p=0$ where $E(0)= \left| \epsilon(0) \right| =\left| \mu \right|$, and is independent of $\bar \Delta_p \Delta_p$. Quite remarkably, this takes us to the previously discussed scenario given by the Eq.~\rf{eq:expa} with ${\bf p}_c=0$. 
Specifically, for $t$ close to  any of the values $t_n$ given by \be \label{eq:tn} t_n=\frac{\pi}{2 \left| \mu \right|} \left( 1 +2n \right), \ee with an arbitrary integer $n$, we can write
\be S_{\bf p} = 2 \cos(t E_p) \approx C \left( t-t_n +  \alpha p^2 \right). 
\ee
Here 
\be C=2 (-1)^{n+1}  \left| \mu \right|,
\ee and 
\be \alpha = \pi \left( \oh + n \right) \frac{\bar \Delta_p \Delta_p m -\mu}{2 \mu^2 |\mu| m }.
\ee Importantly, $\alpha$ is complex due to $\bar \Delta_p \Delta_p$ being complex. Note that this matches the conjectured form \rf{eq:expa}.

Working in the large $V$ limit and replacing summation over ${\bf p}$ with integration we find
\be \label{eq:dzdt}  \frac 1 V \pbyp{ \ln \cZ}{t} = \frac{1}{4\pi} \int  \frac{p \, dp}{t-t_n + \alpha p^2}.
\ee 
The integral above is taken over $p$ varying from $0$ to infinity, although we must remember that only the approximate value for the expression being integrated is written
above valid for  small $p$ only. In particular, that means that the integral above can be cut off at some momentum scale, avoiding any divergencies at large $p$.
It is then straightforward to see that the leading singularity is
\be \frac 1 V \pbyp{ \ln \cZ}{t} \approx -\frac{1}{8 \pi \alpha} \ln \left| t-t_n \right|.
\ee
The expression here is approximate, valid when $t$ is in the vicinity of $t_n$. 

Therefore we arrive at a conclusion advertised earlier. The spectral form factor for the 2D $p$-wave chiral superconductor has periodic logarithmic singularities which occur at times $t_n$, defined
above in Eq.~\rf{eq:tn}. 

It is  important for this argument that $\alpha$ is complex and is not real, which in turn is related to $\bar \Delta_p \Delta_p$ being complex. 

Let us contrast this behavior with that of a conventional $s$-wave superconductor.     Its Bogoliubov-de-Gennes Hamiltonian takes the form
\be \label{eq:ham1} \hat H =  \sum_{\bf p} \left( \begin{matrix} \hat a^\dagger_{\bf p \uparrow} & \hat a_{-{\bf p} \downarrow} \end{matrix} \right) 
\left( \begin{matrix} \epsilon(p) & \Delta_s  \cr \bar \Delta_s & -\epsilon(p) \end{matrix} \right)
\left( \begin{matrix} \hat a_{\bf p \uparrow} \cr \hat a^\dagger_{-\bf p \downarrow} \end{matrix} \right). 
\ee Here $\Delta_s$, $\bar \Delta_s$ are momentum-independent $s$-wave gap functions. 
The spectral form factor takes the same form \rf{eq:sp} but with the spectrum
\be \label{eq:spec} E_s(p) = \sqrt{ \epsilon(p)^2 + \bar \Delta_s \Delta_s}.
\ee
Here $\bar \Delta_s \Delta_s$ is controlled by the gap equation almost identical to the one for the $p$-wave superconductor, given by
\be  \label{eq:gapreal1} \frac  i {2V}  \sum_{\bf p}  \frac{\tan \left[ \frac {t E_s(p)}{2}  \right]} {E_s(p)} = \frac{1}{\lambda}.
\ee
The main point is that, just like in case of Eq.~\rf{eq:gapreal}, the solution of this equation necessarily corresponds to $\bar \Delta_s \Delta_s$ being complex. As a result, $E_s(p)$ is complex. Unlike in case of the $p$-wave superconductor, $E_s(p)$ is complex for all $p$ without exceptions. As a result, none of the factors $S_{\bf p}$ defined in Eq.~\rf{eq:sp} vanish for any time $t$, and the
spectral form factor $\cZ$ is analytic at all times. 

We see that the key distinction between $s$-wave and 2D $p$-wave superconductors was the presence in case of the latter of a
point ${\bf p}=0$ in the gap function $\Delta({\bf p})= (p_x+i p_y) \Delta_p$ where it vanishes. Furthermore, despite having to analytically continue the solution of the gap equation 
\rf{eq:gapim} to  imaginary temperature $1/T \rightarrow it$, we expect that the analytically continued gap function must also vanish as ${\bf p} \rightarrow 0$. Indeed, from the structure of the
Hamiltonian \rf{eq:pwave} the $p$-wave gap function must satisfy
\be \Delta({\bf p}) = - \Delta(-{\bf p}).
\ee
This enforces that the gap function must always vanish at ${\bf p}=0$ even if it is a solution of the analytically continued gap equation \rf{eq:gapreal}. More generally, 
 the key necessarily condition for  a nonanalytic spectral form factor is the gap function $\Delta({\bf p})$ which vanishes at certain values of ${\bf p}$, not only at 
finite temperature, but also when analytically continued to imaginary values of temperature. 

A good second example of a $p$-wave superconductor is class DIII 3D topological superconductor \cite{Ryu2010} (Helium III B phase) with the
the Bogoliubov-de-Gennes Hamiltonian 
\be \label{eq:hamDIII} \hat H = \sum_{\bf p} \left( \begin{matrix} \hat a^\dagger_{\bf p } & \hat a_{-{\bf p} } \end{matrix} \right) 
\left( \begin{matrix} \epsilon(p) & i p_\mu \sigma^y \sigma^\mu \Delta_p)  \cr -i p_\mu  \sigma^\mu \sigma^y \bar \Delta_p & -\epsilon(p) \end{matrix} \right)
\left( \begin{matrix} \hat a_{\bf p} \cr \hat a^\dagger_{-\bf p} \end{matrix} \right),
\ee
where $\sigma^y$ and $\sigma^\mu$ are Pauli matrices acting on the spin indices of the operators  $\hat a_{\bf p}$ and $\hat a^\dagger_{\bf p}$. 
Its spectrum is also given by the equation \rf{eq:pwavesp}, but with ${\bf p}$ being the 3D vector. 
By analogy with the previous analysis leading to equation \rf{eq:dzdt}, we immediately find
 \be \label{eq:dzdt1} \frac 1 V  \pbyp{ \ln \cZ}{t} = \frac{1}{2\pi^2} \int  \frac{p^2 \, dp}{t-t_n + \alpha p^2} \sim \sqrt{\left| t-t_n \right|}.
\ee 

On the other hand, let us examine 2D spin-singlet chiral $d$-wave superconductor, which belongs to the symmetry class C. The corresponding Bogoliubov-de-Gennes Hamiltonian is
\be \label{eq:hamd} \hat H =  \sum_{\bf p} \left( \begin{matrix} \hat a^\dagger_{\bf p \uparrow} & \hat a_{-{\bf p} \downarrow} \end{matrix} \right) 
\left( \begin{matrix} \epsilon(p) & \Delta({\bf p})  \cr  \bar \Delta({\bf p}) & -\epsilon(p) \end{matrix} \right)
\left( \begin{matrix} \hat a_{\bf p \uparrow} \cr \hat a^\dagger_{-\bf p \downarrow} \end{matrix} \right),
\ee
with $\Delta({\bf p}) = (p_x+ip_y)^2 \Delta_d$, $\bar \Delta({\bf p}) = (p_x-ip_y)^2 \bar \Delta_d$. 
What sets this example apart from others is that while the gap function vanishes at ${\bf p}=0$, it is not automatically obvious that the gap function analytically continued
to imaginary temperature would still vanish in this limit. To elucidate this further, we suppose that the gap function consists of both $d$-wave and $s$-wave pieces,
$\Delta({\bf p}) = \Delta_s + (p_x+ip_y)^2 \Delta_d$, $\bar \Delta({\bf p}) = \bar \Delta_s + (p_x-ip_y)^2 \bar \Delta_d$. With rotationally invariant interactions, the gap equation should decouple into two separate equations for $\Delta_s$, $\bar \Delta_s$ and for $\Delta_d$, $\bar \Delta_d$. If $\Delta_s$ is equal to zero for any temperature $T$, its analytic continuation to imaginary values of $T$ should also be zero. At the same time, just as earlier, $\bar \Delta_d \Delta_d$ becomes a complex number, with the spectrum given by $E(p) = \sqrt{\epsilon^2(p) + p^4 \bar \Delta_d \Delta_d }/2$.  leading to the following singularity in the spectral form factor (below $\beta$ is real, while $\alpha$ is complex)
 \be \label{eq:dzdt2} \frac 1 V \pbyp{ \ln \cZ}{t} = \frac{1}{2\pi} \int  \frac{p \, dp}{t-t_n +\beta p^2 + \alpha p^4} \sim \ln \left| t-t_n \right|.
\ee 
However, if $\Delta_s$ is nonzero at some range of temperature, then it may still be nonzero after the analytic continuation $1/(k_B T) \rightarrow it$. Then the superconductor will have a non-singular spectral form factor. To decide whether a particular superconductor of this form will have singularities in its structure factor we need to examine the original 
Hamiltonian of the interacting fermions which led to this superconductor and see if any $s$-wave pairing is possible in addition to the $d$-wave pairing. Therefore, the singularities
in this case are not as ubiquitous as in the $p$-wave case.

All superconductors that we looked at so far were gapped to fermionic excitations. Let us now look at an example of a gapless superconductor. 
As an example, consider a   3D $p$-wave spin-triplet superconductor which has the Bololiubov-de-Gennes equation \rf{eq:ham} with the gap function which behaves as \be \Delta({\bf p}) = \left( p_x+i p_y \right) \Delta_p.\ee 
This gap function vanishes if $p_x=p_y=0$, for all $p_z$. Furthermore, given $\epsilon(p)=p^2/(2m)-\mu$ with $\mu>0$, the excitation spectrum 
\be E({\bf p}) =  \sqrt{ \left(\frac{p^2}{2m}-\mu \right)^2 + \left(p_x^2 + p_y^2 \right) \bar \Delta_p \Delta_p} 
\ee vanishes at $p_x=p_y=0$, $p_z=\sqrt{2 m \mu}$. 

Suppose just as in the previous examples, once the temperature is made imaginary, $\bar \Delta_p \Delta_p$ becomes complex, but otherwise no other terms appear in the gap function. However, unlike the previous examples of gapful superconductors, setting $p_x=p_y=0$, we find that $E(p_z)$ now ranges from zero to infinity. As a result, 
the spectral form factor  ${\cal Z}(t)$ is now an analytic function of time $t$. 

Now it is further possible to imagine that the fermions which formed this  superconductor move on a lattice, as opposed to a continuous space. If so, then  $E(p_z)$, at $p_x=p_0$ now has a maximum somewhere as $p_z$ is varied. Denoting the maximum $E_+$ it is straightforward to see that this would lead to a singularities in the spectral form factor occurring at times $t_n = \pi (2n+1)/(2 E_+)$. These arguments show that singularities are possible even in gapless superconductors, but they are not as ubiquitous and their existence requires some assumptions. 

Note however that if $\mu<0$, then the resulting superconductor is gapful, although not topological. It will still have singularities controlled by $E_- = \left| \mu \right|$. 

Coming back to the gapful (topological) superconductors, we can rely on the classification of the topological superconductors \cite{Ryu2010} to see that there are five distinct classes
of topological superconductors of interest, three in the two dimensional space and two more in the three dimensional space. We can summarize the behavior of their
spectral form factors in the following table.

\begin{center}
\begin{tabular} { |   l  |  l  |  l  | } 
Class & Gap function & Spectral Form Factor \\ \hline
D, $2D$ & $\left( p_x+i p_y \right) \Delta_p$ & $\pbyp{\ln {\cal Z}}{t} \sim \ln \left| t-t_n \right|$ \\
\hline
C, $2D$ & $\left(p_x+ip_y\right)^2 \Delta_d$ & $\pbyp{\ln {\cal Z}}{t} \sim \ln \left| t-t_n \right|$ \\
\hline
DIII, $2D$ & $\left( \sigma^z p_x + i p_y \right) \Delta_p$ & $\pbyp{\ln {\cal Z}}{t} \sim \ln \left| t-t_n \right|$ \\
\hline
DIII, $3D$ & $i p_\mu \sigma^y \sigma^\mu \Delta_p$ & $\pbyp{\ln {\cal Z}}{t} \sim \sqrt{\left| t-t_n \right| }$ \\
\hline
CI, $3D$ & vanishes on surfaces & $\pbyp{\ln {\cal Z}}{t} \sim \sqrt{\left| t-t_n \right|}$ \\
\hline
\end{tabular}
\end{center}
The first three entries in the table above were already worked out above. In particular, class D and class DIII superconductors are $p$-wave and the singularities
in their spectral form factor are ubiquitous. The class C superconductor may have singularities in their spectral form factor if its gap equation excludes the possibility of 
an additional $s$-wave gap function. 
The last entry refers to the class CI topological spin-singlet superconductor in three dimensions 
\cite{Schnyder2009}. It is in the same class as the conventional $s$-wave spin-singlet superconductor and therefore will have singularities in the spectral form factor
only if its gap equation excludes the possibility of an additional $s$-wave gap function. If this is excluded, then working out its singularities relies on the understanding that
its gap function vanishes on 2D surfaces in its 3D Brillouin zone. Starting from the point on the surface where $E(p)$ has its minimum, following the arguments given 
here it is easy to see that
\be 
\pbyp{\ln \cal Z} { t} \sim \int \frac{d^2 q_1 dq_2}{t-t_n-\alpha q_1^2-\beta q_2^2}.
\ee
Here $q_1$ is the coordinate parametrizing the surface and $q_2$ is the direction perpendicular to the surface, $\alpha$ is real while $\beta$ is complex. By analogy with
\rf{eq:dzdt1} we find
\be 
\pbyp{\ln \cal Z} {t}\sim \sqrt{\left| t - t_n \right|},
\ee
just as stated in the table above. 

Therefore, we see that the type of the singularity in the spectral form factor which occurs in topological superconductors depends only on the dimensionality of space. 

Finally, we would like to remark that spectral form factors nowadays are accessible to measuring in experiment, using the techniques of the atomic physics. For example,
if the superconductor is realized by means of cold ions \cite{Shankar2022}, its spectral form factor could in principle be measured by directly evolving a random initial product-state up to some time, projecting it back to the inital state, and averaging over the random initial state. Therefore, spectral form factors can be used as a probe 
of the structure of the superconducting order parameter. 

{\sl Acknowledgement:} We would like to thank E. Yuzbashyan for inspiring discussions. 
VG was supported by the Simons Collaboration on Ultra-Quantum Matter, which is a grant from the Simons Foundation (651440).

\bibliography{library}

\end{document}